\documentclass[fleqn,usenatbib]{mnras}
\usepackage{newtxtext,newtxmath}

\usepackage[T1]{fontenc}
\usepackage{ae,aecompl}

\usepackage{epsfig}
\usepackage{graphicx}
\usepackage{amsmath}
\usepackage{amssymb}
\usepackage{mathrsfs}
\usepackage{mathtools}
\usepackage{hyperref}

\title[Second and third generation disc formation]{The galactic rate of second and third generation disc and planet formation}

\author[Hogg, Wynn \& Nixon]{
Miriam~A~Hogg\thanks{E-mail: mah63@le.ac.uk}, Graham~A~Wynn \& Chris~Nixon\vspace{0.1in}\\
Theoretical Astrophysics Group, Department of Physics and Astronomy, University of Leicester, Leicester, LE1 7RH, UK
}

\date{Draft version, \today.}
\pubyear{2018}
\pagerange{\pageref{firstpage}--\pageref{lastpage}}
\begin{document}
\label{firstpage}
\maketitle

\begin{abstract}
We investigate the formation of discs within binary systems where at least one component has left the main sequence. In particular we calculate the occurrence rates of systems which can host long-lived, massive discs that may be able to support the formation of planets. We synthesize a population based on Milky Way properties, using both theoretical and observational inputs to constrain key properties such as the shape of the initial mass function, binary fraction, and mass transfer physics. We predict 0.26\% of binary systems will host Second generation discs (where the primary has evolved), and 0.13\% of systems will host Third generation discs (where the secondary also evolves). For the Milky Way, this translates into 130 million and 90 million Second and Third generation systems respectively from an estimated total of 50 billion binary systems. Of these systems that form discs, we estimate approximately 20\% of Second and 3.8\% of Third generation discs have enough mass to form a planetary system. We speculate on how the process of planet formation differs in these systems from conventional planet formation in protostellar discs.
\end{abstract}


\section{Introduction}\label{sec:intro}
Binarity and multiplicity are common features in star systems. Around 50\% of Sun-like G stars are in binaries and that percentage increases to 70\% in A stars and up to 100\% in O stars \citep[e.g.][]{duquennoy_mayor_1991,raghavan_mcalister_2010,duchene_kraus_2013}. Planets have been observed around both single and binary stars. While several thousand planets have been found to date, only a few 10s have been found in binary systems. Most of the planets in binary systems are in S-type orbits (inside the binary, orbiting one of the two stars), while the rest are in P-type orbits (outside of the binary, orbiting both stars).

Current understanding of the formation of planetary systems can typically be divided into two main theories: 
1. core accretion where small particles and planetesimals coagulate growing into planets which can later accrete gas \citep[e.g.][]{mizuno_1980, lissauer_1993, pollack_hubickyj_1996} and 
2. gravitational instability where the protoplanetary disc fragments in gaseous clumps which condense into planets \citep[e.g.][]{adams_ruden_1989, boss_1997,durisen_boss_2007}. 

Both of these theories are conventionally used in the context of protoplanetary discs where the planets form from the remnant material left over from the formation of the central star. After a few million years a combination of accretion and photoevaporation removes the disc and conventional planet formation ends. 

It is possible that planets may be able to form much later in a star's lifetime.  In particular, AGB or post-AGB phase (planetary nebulae) stars suffer significant mass loss, which, when the star is in a binary system, may be caputered by its binary companion. In some cases, this captured mass may form a substantial, perhaps protoplanetary, disc. Discs formed during the evolution of the more massive binary component are known as `Second generation discs' and can be circumprimary, circumsecondary or circumbinary. Discs formed later in the evolution of the binary system, during the late evolutionary stages of the (originally) lower mass component, are termed `Third generation discs' \citep{peret_2010} and can also form around the primary, secondary or binary itself. 

Observationally it is uncertain whether any planets formed in a Second or Third generation discs have been found. However, the planets around pulsar PSR B1257+12 are orbiting too close to have survived the giant branch of the star and with an eccentricity low enough that their presence is difficult to explain if they are First generation planets \citep{martin_livio_2016}. These planets may have formed from a `fallback disc' of material that was unable to escape the system.

The properties of second and third generation discs are uncertain, as they depend on the nature of the mass loss. The speed and dust species in the ejecta affect the subsequent dynamics and these quantities depend on the stellar properties. However the general properties of mass-transfer between the stars in binary systems are reasonably well-constrained by theoretical and observational studies. The distance between the stars determines the nature of the mass transfer. If the stars are widely separated $>200$\,AU they will evolve nearly independently and have little effect on each other. If the stars are in a tight orbit $<5$\,AU they will interact strongly with multiple mass transfer events, some even before they leave the main sequence.  Intermediate separations ($5-200$\,AU) involve mass transfer at rates that do not cause drastic changes in stellar evolution. It is these less intense mass transfer phases that cause chemical changes in stellar atmospheres, such as barium enrichment \citep{mcclure_fletcher_1980}. 
 
Simulations of mass transfer between evolving binaries have been conducted in two main categories: wind mass transfer where the accreting star captures material from the wind of the donor star and Roche lobe overflow where the donor star fills its Roche lobe and transfers material through the inner Langrange (L1) point. A third, less-studied, transfer process occurs when the donor star gives off a slow wind which is captured predominantly near the L1 point and is called wind Roche lobe overflow \citep{mohamed_podsiadlowski_2007}. Wind mass transfer occurs primarily in wider binaries and is described by the Bondi-Hoyle equations \citep{bondi_hoyle_1944}. Some studies have focused on the specific aspects of wind accretion such as mass ratio \citep[e.g.][]{nagae_matsuda_2004,jahanara_mitsumoto_2005}, and the effect the companion has on the donor by increasing the efficiency of mass transfer via wind focusing \citep[e.g.][]{skopal_carikova_2015,shagatova_skopal_2016, val-borro_karovaska_2009}. There has also been a comparative study of the observations to the theory \citep{shakura_postnov_2017}. For Roche lobe overflow, the dynamics of mass-transfer through the L1 point has been studied in both circular and eccentric binaries using smoothed particle hydrodynamics (SPH) simulations \citep[e.g.][]{lajoie_sills_2011,chuch_dischler_2009,staff_demarco_2014}. There have also been studies of non uniform mass accretion \citep{gharmi_ghosh_2014}, mass escape through the L2 point \citep{linial_sari_2017}, orbital evolution \citep{dosopoulou_kalogera_2016} and stability \citep{negu_tessema_2015}. We discuss the mass transfer processes in more detail in Section~\ref{sec:method}.

After the main-sequence, stars of mass less than $\sim 8M_\odot$ will reach the asymptotic giant branch (AGB) after going through the red giant and horizontal branch phases. During the AGB phase there is significant mass loss through winds and pulsations. The AGB phase lasts a relatively short time (approx 5\,Myr) so there are expected to be a few hundred stars in the Galaxy currently in the AGB or post-AGB phase \citep{belczynski_mikolajewska_2000}. There have been several observations of circumsteller discs around AGB and post-AGB stars. \cite{bujarraball_castro-carrizo_2015} report interferometric data of the $^{12}$CO $J=2-1$ emission, revealing a Keplerian disc around a post-AGB star. A similar study by \cite{hillen_van-winckel_2017} used an interferometric IR survey to compare signatures of post AGB binaries to young stellar objects (YSO) to look for signs of disc formation. They found that discs that formed in binary systems with an AGB component are comparable to YSO discs. \cite{van-winckel_lloyd-evans_2009} used radial velocity and photometric data to discern the presence of circumbinary discs in six post-AGB systems. Circumsecondary discs have been found by long-term studies of symbiotic stars \citep[e.g.][]{van-winckel_2017} and infrared data has been used to infer a disc around the companion of the symbiotic binary Mira AB \citep{ireland_monnier_2007}.  

\cite{perets_kenyon_2013} considered discs formed via mass transfer as a site of potential `Second generation' planet formation. They focused on wind-fed discs, arguing that discs from Roche lobe overflow were too luminous and short-lived to allow planet formation -- in contrast discs formed from wind capture were cooler and longer lived. \cite{perets_kenyon_2013} used a combination of analytical and numerical approaches to model the long-term evolution of these discs. They focused on the AGB phase of the evolution as this provides the necessary mass-loss rates to accumulate a useful amount of mass in a relatively short time. They found that discs formed in binaries with separations between $3-100$\,AU in 16 models with different mass-loss rates, indicating that Second generation planets may be possible if they can form within such discs. Previous work by the same author cites Gl-86 as a system that may contain a second generation planet, due to the orbital configuration. (see \cite{peret_2011} for further details)

In this paper we seek to estimate the formation rates of Second and Third generation discs. We develop a synthetic population of binaries and use analytic estimates of the binary and stellar properties to determine if mass transfer occurs and whether discs form. We include discs formed via winds in the Bondi-Hoyle regime and those formed via wind Roche lobe overflow \citep{mohamed_podsiadlowski_2007}. Using both of these formation mechanisms we can explore a large range of both intermediate and wide separation binaries. Following \cite{perets_kenyon_2013} we exclude the close binaries that would undergo standard Roche lobe overflow. In Section~\ref{sec:modelpop} we described our population synthesis methodology and parameter choices. In Section~\ref{sec:method} we describe the mass transfer processes and their properties and how we calculate the presence and properties of the discs. In Section~\ref{sec:results} we present our results. Finally in Sections~\ref{sec:discussion} \& \ref{sec:con} we provide discussion and conclusions.

\section{Population synthesis}\label{sec:modelpop}
\subsection{Overview}\label{sec:overview}
We use observational and theoretical constraints to synthesise the population of binary systems expected in the Milky Way. In particular, we use observation based estimates of the distributions of stellar and binary parameters to establish an initial population and implement analytic prescriptions for stellar and binary evolution to identify the systems that undergo mass transfer. We are then able to scan parameter space to identify trends in mass transfer and disc formation. We use our results to estimate the number of systems that have discs, and infer from their properties (mass and radii) their potential for forming planetary systems. \\

We create systems with primaries (higher mass component) with mass $M_{1}$ and secondaries (lower mass component) with mass $M_{2}$, orbital periods $P$, eccentricities $e$, and ages $t$, chosen to reflect Milky Way values. We use theoretical evolution equations to find the radius, luminosity, wind speed and final mass of the star when it enters the giant phase, specifically the asymptotic giant branch (AGB) where the star experiences large rates of mass loss. Using these values we calculate whether the mass lost from the primary would transfer to the secondary via Bondi-Hoyle stellar wind, Wind-Roche lobe overflow or Roche lobe overflow. \\

We focus on the AGB phase as it is a period of high mass-loss which would be more likely to cause disc formation. Stars that have an AGB phase will fall in the mass range of $0.8-7M_\odot$. Stars up to $8M_\odot$ may have an AGB phase but are categorised as a `super-AGB', which do not follow the same mass-loss mechanisms and lifetimes as normal AGB phases, so we do not include these.

\subsection{Parameters}\label{sec:params}
\subsubsection{Primary mass}\label{sec:mass}
We draw the primary star's mass from an initial mass function (IMF). The shape of the IMF dictates the number of primary stars that will be in the mass range of interest. There are several IMFs in the literature which we can choose from. The \cite{salpeter_1954} IMF uses a linear relationship for the entire mass range. This leads to more low-mass stars than the \cite{kroupa_2001} IMF which uses a broken power law or the \cite{chabrier_2005} IMF which has a smoother curve using an exponential. The number of primary stars below $0.7 M_{\odot}$ varies by nearly 10\% among the three IMFs. For the Salpeter IMF 92.1\% of stars are below $0.7 M_{\odot}$, leaving only 7.9\% of stars in the desired range. The Kroupa IMF has 88\% of systems under $0.7 M_{\odot}$ which leaves a few more in the range of focus. The Chabrier IMF has 85\% under the $0.7 M_{\odot}$ range leaving 15\% that can be used. The difference between the Salpeter and Chabrier IMF is a factor of 3 in the number of stars under the $0.7 M_{\odot}$ limit, which should be taken into consideration when interpreting our results, for which we employ the \cite{chabrier_2005} IMF. The upper limit of stars we are interested in is the highest mass that will have a normal AGB phase without supernova or other outburst events and will end their lives as white dwarfs. Thus we set an upper limit of $7 M_{\odot}$. Stars with mass as low as $0.6M_\odot$ can go through an AGB phase. However, we set the lower limit at $0.7 M_{\odot}$, as this is the lowest stellar mass that can evolve to this phase in the age of the galaxy \citep{Hansen_1999}.

\subsubsection{Mass ratio}\label{sec:massratio}
The mass ratio distribution is uncertain with several variations in the literature. We have explored several of these \citep[e.g.][]{raghavan_mcalister_2010,reggiani_meyer_2013,duchene_kraus_2013,gullikson_kraus_2016}, finding that the fraction of systems hosting discs at each stage differs by less than a factor of two. This difference is mostly caused by the propensity of discs to form within binaries that have near-equal mass ratios (see Fig.~\ref{fig:9} below).

For the remainder of the paper we use the results of \cite{duchene_kraus_2013}. They collate data from previous observational studies to find the multiplicity, mass ratio, and orbital separation for nearby stars. Specifically they separate the stars into groups based on primary mass and find each have different values. The study looks at primary masses in the entire mass range: $0.1M_{\odot}<M_1<8M_{\odot}$. We are interested in stellar masses with $0.7M_{\odot}<M_1<7M_{\odot}$ which Duch\^{e}ne \& Kraus label as either `solar type' ($0.7 M_{\odot}<M_1<1.7M_{\odot}$) or `intermediate mass' ($1.5<M_{\odot}<5M_{\odot}$). For solar type stars the mass ratio $q=M_{2}/M_{1}$ is split into two power laws depending on the period, $P$ in days, given by
\begin{equation}
  \gamma_{0.7-1.5}^{logP<5.5} = 1.16 \pm 0.16 \\
  \gamma_{0.7-1.5}^{logP>5.5} = -0.01 \pm 0.03\,.
\end{equation}
For intermediate mass stars the distribution has no simple analytical representation due to incompleteness of the data and selection biases, so for this we  assume \citep{duchene_kraus_2013}
\begin{equation}
  \gamma_{1.5-5} = -0.45 \pm 0.15\,.
\end{equation}

\subsubsection{Eccentricity}\label{sec:ecc}
\cite{tokovinin_kiyaeva_2016} suggest a Gaussian eccentricity distribution with a peak at 0.59 and $\sigma$ of 0.25. In contrast \cite{duchene_kraus_2013} favour a flat distribution. We have tested our results with both distributions and find the difference in the number of discs created to be at the $1-10$\% level. The higher eccentricity systems are more likely to create discs due to episodic higher mass transfer around periastron that would not occur in less eccentric systems. The results we present in this paper use the \cite{tokovinin_kiyaeva_2016} distribution.

\subsubsection{Orbital period}\label{sec:orbit}
For the binary orbital period we use a log-normal curve with a mean of $10^{4.3}$ days and a $\sigma$ of $10^{2.3}$ \citep{duquennoy_mayor_1991}. This distribution creates some systems with $>10^{9}$ day orbits which is too far apart to interact, and it also creates some very short period systems. We impose a cutoff in binary separation at $3000$\,AU as beyond this value galactic tides will begin to affect the binary evolution \citep{bonsor_veras_2015}. Mass transfer in binary systems with separations near this limit is likely to be insignificant, but we include them for completeness.

At small separations, $\lesssim 0.1$\,AU, Roche lobe overflow is likely to occur, and thus no planet forming disc is expected \citep{perets_kenyon_2013}. We return to this in more detail in Section~\ref{sec:method}. The minimum separation for wind Roche lobe overflow or wind-fed discs to be planet forming is set by the distance from the AGB (and later the WD) star. \cite{marzari_thebault_2007} found that planets can form in binaries with separations around $20$\,AU if the eccentricity is low. \cite{tutukov_fedorova_2012} find that there are no S-type planets observed in binary systems with a separation of $10$\,AU or lower. However, it is unclear if this picture will change with future observations, or whether planet formation in Second and Third generation discs proceeds differently. As we have noted, planets appear to have formed at small radii in pulsar fallback discs.

\subsubsection{Stellar age}\label{sec:age}
\begin{figure}
  \begin{center}
    \includegraphics[width=\columnwidth]{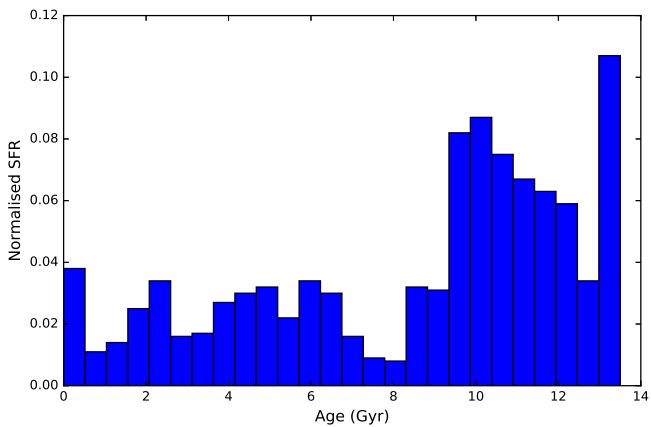}
    \caption{The normalised star formation rate (SFR) of the Milky Way \protect\citep{snaith_haywood_2015}.}
    \label{fig:1}
  \end{center}
\end{figure}

In order to predict the current number of systems in the Milky Way with Second or Third generation discs we need to look at the star formation history of the Galaxy. We use these to find a distribution of ages for the stars in our population synthesis. The star formation history of the Milky Way cannot be easily modelled by a curve or function. The Milky Way had a few bursts of star formation, one between 9-13 Gyr ago and one between 2-6 Gyrs ago. The first star formation burst produced over half of the Galactic stellar mass. We used a model for the star formation history of the Milky Way from \cite{snaith_haywood_2015} to give each system an age that matches the Milky Way distribution (see Fig.~\ref{fig:1}).

\subsubsection{Parameter correlations}\label{sec:param_cor}
A study by \cite{moe_di-stefano_2017} explored correlations between binary parameters such as period, mass ratio, and eccentricity. Like \cite{duchene_kraus_2013}, which we employ in our code, they find a power law correlation between period and mass ratio, although the values are different. We find that the effect of changing the period and mass ratio correlation changes the number of discs formed by a factor of 2 or 3. Eccentricity effects the likelihood of disc formation more than the other parameters and for wide binaries, which are our focus here, the results of \cite{moe_di-stefano_2017} agree with the Gaussian distribution found by \cite{tokovinin_kiyaeva_2016}, used in our study.

\section{Mass transfer}\label{sec:method}
\subsection{Regimes}
There are three main types of mass transfer in post main sequence binaries: Roche lobe overflow (hereafter RLOF), wind Roche lobe overflow (WRLOF) and Bondi-Hoyle stellar wind (BHW).

RLOF occurs in close binary systems. When one of the stars overfills its Roche lobe the material flows into the potential well of the second star through the L1 point. This normally occurs when the star expands or loses mass causing its Roche lobe to shrink. RLOF normally occurs post main sequence when the star travels up the giant branch. The material from the donor star flows through the L1 point to the secondary so the mass is rarely unbound from the system and the secondary can accrete up to 100\% of the transferred mass. RLOF can be either stable or unstable depending on how the mass losing star reacts to the mass loss. If the star's radius shrinks in response to the mass loss the mass transfer can be shut-off and is stable. However if the star expands in response to mass loss, which often happens on the giant branch due to the convective envelope, the mass transfer becomes unstable and causes a common envelope and inspiral phase \citep{pacyzynski_1971}.

BHW occurs in wide binaries where the donor star loses mass through a wind on the giant branch. The wind is assumed to be spherical and the secondary will travel through some of this material and gather it up into an accretion disc. This mechanism has a small capture fraction of a few per cent \citep{bondi_hoyle_1944}.

WRLOF is similar to RLOF, but instead of the star itself overflowing the Roche lobe, a slow wind from the giant star fills the Roche lobe. This causes matter to be focused towards the L1 point as in the RLOF mechanism and gives a higher rate of capture onto the secondary than BHW, up to 70\%. For this to occur the star has to be filling a large enough fraction of the Roche lobe that the slow wind can form from dust sublimation in the upper atmosphere of the giant donor. Dust sublimation in this way mainly occurs in the asymptotic giant branch so the star has to be in this phase for this mass transfer regime to work \citep{mohamed_podsiadlowski_2007}.

The WRLOF mechanism was suggested based on in depth studies of Mira AB \citep{podsiadlowski_mohamed_2007,ramstedt_mohamed_2014}. Similarly, a study on SS Leporis \citep{blind_boffin_2011} also finds that the WRLOF regime explains a lot of the observed properties of the system, and some that cannot be explained by the BHW mechanism alone. \cite{abate_pols_2013} used the WRLOF regime to explain the discrepancy between the observed carbon enhanced metal poor (CEMP) stars and theoretical predictions. There have been no studies of WRLOF and its impact on Second and Third generation planet formation. \cite{perets_kenyon_2013} looked at wind-formed discs in binaries, however they focused only on BHW as the mass transfer mechanism. We use a combination of WRLOF and BHW to explore mass transfer in binary systems and whether discs are formed. We do not include RLOF as the mass transfer from a giant donor star is normally unstable, and the resulting discs are too small and hot to be considered for planet formation. The mass transfer regime each individual systems falls into depends on how much of the Roche lobe is being filled by the donor star.

\subsection{Roche lobe calculation}\label{sec:rochelobe}
The Roche lobe size determines which mass transfer regime a system will be in and therefore its calculated capture fraction and disc mass. Most wide binary systems are non synchronous and eccentric so the Roche lobe formalism of \cite{eggleton_1983} cannot be reliably used in most circumstances. \cite{spinsky_willems_2007} describe a method that includes the non-synchronous eccentric nature of the binaries and gave a more accurate value for the Lagrange points. The potential surfaces of the Lagrangian points can be very different from the standard Eggleton formula depending on the eccentricity and spin and when calculating mass loss these are important considerations.

Here we need the Roche lobe L1 point for a binary system composed of a giant star and its companion. Following \cite{spinsky_willems_2007} we begin with the ratio of the rotational angular frequency, $\Omega_{\rm AGB}$ and orbital angular frequency $\omega_p$,
\begin{equation}\label{eq:f}
  f=\frac{|\Omega_{\rm AGB}|}{\omega_{\rm p}}\,.
\end{equation}
Note that for positive binary frequency $\omega_{\rm p}$, the stellar rotational frequency can be positive (prograde) or negative (retrograde).

The orbital angular frequency of the system can be calculated as
\begin{equation}\label{eq:wp}
  \omega_{\rm p}=\frac{2\pi}{P} \frac{(1+e)^{1/2}}{(1-e)^{3/2}}\,,
\end{equation}
for a given orbital period, $P$, and eccentricity, $e$.

We can approximate $\Omega_{\rm AGB}$ by assuming the angular momentum of the giant is conserved from when it was a main sequence star, and thus
\begin{equation}\label{eq:omegafinal}
  \Omega_{\rm AGB} = \frac{R_{\rm MS}V_{\rm MS}}{R_{\rm AGB}^{2}}
\end{equation}
where $R_{\rm AGB}$ is the radius of the star during the AGB phase, $R_{\rm MS}$ is radius of the star on the main sequence (of mass $M_{\rm MS}$), and the rotation velocity $V_{\rm MS}$ of the main sequence progenitor is \citep{hurley_pols_2000}
\begin{equation}
V_{\rm MS}=\frac{330M_{\rm MS}^{3.3}}{15+M_{\rm MS}^{3.45}}~~{\rm km/s}\,.
\end{equation}

Sepinsky et al's method requires the value of $f$ to be of order unity to be considered a `quasi-static' system and thus to provide an accurate result. Then equations (21) and (25) of \cite{spinsky_willems_2007} are used to calculate the L1 point in these systems. Their motivation was to interpret observations of eccentric and non synchronous close systems that could undergo Roche lobe overflow. In these fairly close systems the value of $f$ is often close enough to unity for the quasi-static approximation to hold. However we are also considering wider separation systems where mass transfer can occur in the WRLOF regime in a similar fashion to RLOF. Due to this increased separation, $f$ may not always be close to unity. In these cases we cannot use the Sepinsky et al. method, and instead revert to the \cite{eggleton_1983} formula
\begin{equation}\label{eq:ERLO}
  R_{L1}=D(t)\frac{0.49q^{2/3}}{0.6q^{2/3}+\ln(1+q^{1/3})} 
\end{equation}
where $D(t)$ is the instantaneous distance and $q$ is the mass ratio $q=M_{2}/M_{1}$. The distance is given by
\begin{equation}
  D(t)=\frac{a(1-e^{2})}{1+e\cos\nu}
\end{equation}
where $\nu$ is the true anomaly and $e$ is the eccentricity of the binary

We then use the radius of L1 point to calculate its ratio to the dust sublimation radius, $R_{\rm dust}$, given by
\begin{equation}\label{eq:x}
  x=\frac{R_{\rm dust}}{R_{L1}}\,.
\end{equation}
$R_{\rm dust}$ is the distance from the giant star where the dust sublimates and a slow wind can form and is estimated as \citep{abate_pols_2013}
\begin{equation}\label{eq:dustsub}
  R_{\rm dust}=3R_{\rm AGB}\,.
\end{equation}
The fraction $x$ can be used to determine which of the three mass transfer mechanisms is operating in the binary system. 

\subsection{AGB radius}
We assume that the giant is in the AGB phase as it is the phase of most mass loss and has the conditions necessary for the slow wind to form via dust sublimation. To calculate the radius of the donor star in this phase, we first calculate its final mass using \citep{catalan_isern_2008}
\begin{equation}\label{eq:IFmass}
  \begin{array}{l}
 {\rm For}~M_{\rm initial}<2.7M_\odot{\rm :}~~~M_{\rm final} = 0.096M_{\rm initial}/M_{\odot}+0.429\\
 {\rm For}~M_{\rm initial}>2.7M_\odot{\rm :}~~~M_{\rm final} = 0.137M_{\rm initial}/M_{\odot}+0.318
 \end{array}
\end{equation}  

We also require the AGB star luminosity. It is known that most AGB stars have a `superwind' phase at the end of the AGB where the envelope is shed so the wind speed and mass loss increases around this time. For simplicity we do not include this phase. The equation for their luminosity differs for different core masses and metallicities. However, the Paczynski relation \citep{pacyzynski_1971} assumes solar metallicity and is valid for $0.52<M_{\rm final}<0.7$ which is where most of the primary stars are expected to end up \citep{boothroyd_sackman_1988}. Thus we employ this relation here, given by
\begin{equation}
  L = 52,000\left(\frac{M_{\rm final}}{M_{\odot}}-0.456\right)L_\odot\,.
\end{equation}

With the AGB star mass and luminosity, we can calculate the radius following \cite{hurley_pols_2000} who give equations for the post main-sequence evolution of a star, from the first giant branch to the white dwarf. This equation gives values which match well with observed AGB stars. The radius equation is
\begin{equation}
  R_{\rm AGB}=1.125\left(\frac{M_{1}}{M_{\odot}}\right)^{-0.33}\left[\left(\frac{L}{L_{\odot}}\right)^{0.4}+0.383\left(\frac{L}{L_{\odot}}\right)^{0.76} \right]
\end{equation}

While there are equations to find the exact dust sublimation radius based on temperature, most are found to be in the range of ~3 times this radius. So we approximate the dust sublimation radius as (eq.~\ref{eq:dustsub}).

\subsection{Mass transfer equations}\label{sec:MTE}
We use the following mass transfer equations for both Second and Third generation systems. The cutoff between the three regimes are supplied by \cite{abate_pols_2013} as well as the equations for the BHW and WRLOF. First we find how much of the Roche lobe is filled by the donor star following Section~\ref{sec:rochelobe}. Then if $x<0.4$ BHW occurs, if $0.4\le x<3$ WRLOF occurs, and if $x\geq 3$ RLOF occurs.

We note that during the AGB, the donor star loses mass causing the binary orbit to evolve. We calculate that the estimate of the Roche lobe radius changes (increases) by a factor of 2-3 by the end of the AGB phase. Thus the estimate of the capture fraction in each regime is changed by a small factor. We therefore calculate the capture fraction by averaging its value (given below) between the start and end of the AGB phase. We anticipate that the error induced by this is smaller than the uncertainty in the capture fractions for any given system.

\subsubsection{Bondi-Hoyle}
The Bondi-Hoyle capture fraction is given by 
\begin{equation} \label{eq:bh}
  CF_{\rm BHW}=\frac{\alpha}{2\sqrt{1-e^{2}}}q^{2}\left(\frac{GM_{1}}{a v_{w}^{2}}\right)^{2} \left[1+(1+q)\frac{GM_{1}}{a v_{w}^{2}}\right]^{-3/2}
\end{equation}
where $\alpha=1.5$ is a constant, and $v_{w}$ is the wind speed. The wind velocity is difficult to observe but for most giant-star winds, particularly those in AGB stars, they are in the range of $5-35$\,km/s. Following \cite{hurley_tout_2002} we take the wind velocity as
\begin{equation}
  V_{w}=2\beta_{W}\sqrt[]{\frac{GM_{1}}{R_{1}}}
\end{equation}
In the case of cool giant stars, $\beta_{W} \approx 1/8$. 

\subsubsection{Wind-Roche Lobe Overflow}
The WRLOF capture fraction is given by \citep{abate_pols_2013}
\begin{equation} \label{eq:wrlof}
  CF_{\rm WRLOF}=\frac{25}{9}q^{2}\left[-0.284x^{2}+0.918x-0.234 \right]
\end{equation}
This was determined by fitting to 5 simulation models of a $1M_{\odot}$ primary and $0.6M_{\odot}$ secondary at varying separations. \cite{abate_pols_2013} caution that they did not explore the effects of varying the mass ratio, and instead assume the functional form of the BHW value, i.e. $q^2$. They suspect that the true dependence on mass ratio is weaker. The equation also does not include the effect of eccentricity, so to give a more realistic value for this case, we average the capture fraction around the orbit.

\subsubsection{Roche lobe overflow}\label{sec:rlo}
Roche lobe overflow has high mass transfer rates. However mass transfer from a giant star with a convective envelope is thought to be unstable due to the hydrostatic equilibrium condition expanding the radius of the star in response to mass loss \citep{pacyzynski_1971}. The same mass loss also shrinks the Roche lobe of the mass-losing star. The timescale of the radius increase is of the order of the pulsation timescale and this results in unstable mass transfer. AGB stars have convective outer envelopes so it is likely that any systems with RLOF where the AGB star is the donor would be unstable and result in a common envelope phase. The common envelope phase can end in either a merger or a close orbiting binary system. We assume that systems where RLOF and common envelope occurs will not end in a Second or Third generation disc that could be the site of planet formation, so they can be ignored for our purposes. 

\subsubsection{Moving between regimes}
The addition of eccentricity in our calculations means the Roche lobe changes over the orbit of the star and can therefore move between different regimes over an orbit. To find the average capture fraction per orbit we split the orbit up into 360 chunks and find the $x$ value at each point, from the $x$ value we can find which mass transfer regime and calculate the capture fraction at that point which we then use to find the average. If the orbit ever moves into the Roche lobe overflow regime it it is assumed that a common envelope phase results.

\subsection{Second generation disc calculations}\label{sec:SGD}
A Second generation system for our purposes is composed of an AGB star and a main sequence companion. We focus our attention on those systems where mass transfer is stable. At the start of each calculation we check the age of the system; each system is given an age based on the star formation history of the Milky Way (Section \ref{sec:age}). The age determines the star's current properties, in particular whether it has evolved beyond the main sequence. We also compare this to the main sequence turn-off time of the companion. If the two stars are too close in age the companion will reach the AGB at approximately the same time as the primary and a Second generation disc would not be able to form around it.

If the system is in the correct age range for the primary to evolve and is at a separation where mass transfer occurs, we calculate the capture fraction and then apply a criterion for whether the disc is able to form. These criteria are different depending on the mass transfer regime

For BHW the disc forms from the accretion of material from the stellar wind as the secondary orbits the primary. Thus for a disc to form the material accreted from the wind must have more angular momentum, in the frame of the secondary, than the angular momentum of an orbit at the equator of the secondary.  This condition is given by \cite{soker_2004}:

\begin{multline} \label{eq:diskformBH}
  0.25\left(\frac{\eta}{0.1}\right)\left(\frac{M}{2.5M_{\odot}}\right)^{1/2} \left(\frac{M_{2}}{0.6M_{\odot}}\right)^{3/2}\left(\frac{R_{2}}{R_{\odot}}\right)^{-1/2} \\\ \times\left(\frac{a}{100\,{\rm AU}}\right)^{-3/2}\left(\frac{v_{w}}{10\,{\rm kms}^{-1}}\right)^{-4}>1
\end{multline}
where $M=M_1+M_2$ is the total binary mass and $\eta$ is a parameter that is $\sim 0.1$ for isothermal gas accretion and $\sim 0.3$ for adiabatic accretion.  

For WRLOF the transferred matter streams through the L1 point so it is already within the gravitational potential of the companion, in this case the formation of the disc is contingent on if the matter stream impacts the companion. The criterion is given by the circularisation radius being larger than the accreting stellar size \citep{king_raine_1992}:
\begin{equation}\label{eq:diskformWRLOF}
  R_{\rm circ}/R_{\odot} = 2(1+q)^{4/3}[0.500-0.227\log(q)]^{4} P_{\rm days}^{2/3}.
\end{equation}
If $R_{\rm circ}$ is less than the stellar radius then the matter directly impacts the star and a disc cannot form. We calculate the value of $R_{\rm circ}$ for all of the systems and we find that the value is always above $20R_\odot$, so we assume that the discs always form in this case. 

In systems that move between two regimes we employ the criterion for the higher mass transfer rate mechanism (i.e. the WRLOF). For the systems that can form a disc, we then calculate an estimate of the disc mass as:
\begin{equation}\label{eq:discmass}
  M_{\rm disc} = {\rm CF} (M_{\rm initial}-M_{\rm final}),
\end{equation}
Where the value of ${\rm CF}$ is found from either (\ref{eq:bh}) or (\ref{eq:wrlof}) and the final mass of the primary is found using the equations give in (\ref{eq:IFmass}).

To approximate the maximum outer radius of the S-type discs we used the tidal truncation radius from \cite{pichardo_sparke_2005}. To calculate this, they used particles to trace out closed loops that changed periodically due to the eccentric binary. These `Invariant loops' were used to find the outermost stable orbits for S-type discs and innermost stable orbits for circumbinary discs. They found the outer disc radius can be estimated as
\begin{equation}\label{eq:diskrad}
  R_{\rm disc} = 0.733R_{L1}(1-e)^{1.20}q^{0.07}\,.
\end{equation}
\cite{pichardo_sparke_2005} calculate that this is accurate to within $6.5$\% and find good agreement with discs in $\alpha$ Centauri and L1551.

\subsection{Third generation disc calculations}\label{sec:TGD}
After the primary's AGB phase is finished and it is cooling into a white dwarf, we turn our attention to the secondary. The system now consists of a white dwarf that was once the primary and the main sequence companion that may now be the more massive component. The main sequence star will also evolve through its giant phase and can transfer mass back to the original primary so the same calculations need to be done to check for Third generation discs. 

The mass loss from the primary in its AGB phase causes the orbit to widen and we use this new orbit in the Third generation calculations. We assume for simplicity that little mass is accreted onto the companion, so mass lost from the AGB can be assumed to have been lost from the entire system. This amounts to assuming that photoevapouration is the dominant loss of Second generation disc material, rather than accretion on to the secondary star. The loss of mass and angular momentum from the system would widen the orbit to
\begin{equation} \label{eq:newA}
  a_{\rm new} = \left(\frac{M_{1}+M_{2}}{M_{1}+M_{2}-\Delta M_{1}}\right)a_{\rm old}\,.
\end{equation}
Due to the assumption that all of the transferred mass is lost from the system, this new distance is the maximum separation that could occur.  

The Third generation checks are similar to section \ref{sec:SGD}. The maximum and minimum separation of 3000AU and 20AU remain the same. We again check the companion age to see whether it would have evolved off the main sequence within the current age of the Milky Way. We then employ the same disc formation conditions as Section~\ref{sec:SGD} with the new binary parameters.

\subsection{Decision Tree}
Below we show the decision tree the code uses to employ the correct equations in each individual system:
\begin{enumerate}
\item \textbf{Find $f$ value  (Equation \ref{eq:f})}
\begin{itemize}
\item If $f > 4$ use the Eggleton formula to calculate Roche lobe radius (Equation \ref{eq:ERLO})
\item if $f < 4$ use the sepinsky formula to calculate Roche lobe radius (Equation 21 and 25 from \citep{spinsky_willems_2007})
\end{itemize}
\item \textbf{Find $x$ value  (Equation \ref{eq:x})}
\begin{itemize}
\item if $x >3$ Roche lobe overflow occurs: system discarded
\item if $0.4<x<3$ wind roche lobe overflow regime used to find capture fraction (Equation \ref{eq:wrlof})
\item $x<0.4$ Bondi-Hoyle wind regime used to find capture fraction (Equation \ref{eq:bh})
\end{itemize}
\item \textbf{Calculate if disc forms:}
\begin{itemize}
\item if capture is in BHW regime, use equation \ref{eq:diskformBH} 
\item if capture if in WRLOF regime, use equation \ref{eq:diskformWRLOF} 
\end{itemize}
\item \textbf{Derive final mass of disc (Equation \ref{eq:discmass})}
\end{enumerate} 
We follow this method in both Second and Third generation systems. It is possible for a binary system to have a Second generation accretion disc but not a Third generation one, and vice versa, both or neither. As we can have any combination of the above we keep all of the systems for the evolution of both stars.

\section{Results}\label{sec:results}
Following the methodology described above, we perform a Monte-Carlo population synthesis to produce a set of binary systems expected for a Milky Way galaxy. To try and create an accurate representation of the Milky Way our synthesis includes those systems that do not undergo an AGB phase or mass transfer which we keep to track the overall number of systems with and without second generation discs and the characteristics of all systems. The number of stars in the Milky Way is thought to be between 200-400 billion. Taking into account the multiplicity fractions as a function of stellar mass, and the IMF \citep[e.g.][]{duchene_kraus_2013}, we estimate that (assuming 50\% of systems are singles, 30\% are binaries, 20\% are triples, and assuming higher multiples do not significantly contribute) there are 60-120 billion single stars, 35-70 billion binaries and 25-50 billion triple systems. It is infeasible to produce a population synthesis this large, and to understand the large scale trends we are interested in we do not need to, so instead we produce a population of 10 million binary systems and scale our results appropriately. Where we report numbers of systems below, we scale our results to the expected number of binary systems, which we take as 50 billion.

From our analysis we find that a small, but non-negligible, percentage of systems form discs that may be viable sites for planet formation. The detailed numbers from our population are given in Table~\ref{table:results}. The majority of systems do not form or have not yet formed discs in the lifetime of the Milky Way. Of these, most have masses $<0.7M_\odot$ below which they will not or have not yet reached the AGB phase, and the remainder are e.g. in the RLOF regime, or have separations that are too large or too small to host planet-forming discs. Of the systems that can form potentially viable discs for planet formation (0.26\% at Second generation phase, and 0.18\% at Third generation phase), we find that less than 10\% of these systems are exclusively either BHW or WRLOF systems, while the rest exhibit both. We find that systems with a high mass ratio or high eccentricity have a higher likelihood of significant mass transfer. We also find that systems with a high mass primary have a much higher chance of forming discs due to their AGB radius being larger and thus they are more likely to be in the higher mass transfer (WRLOF) regime.

\begin{table}
\centering
  \caption{The outcome of our population synthesis of binary systems in the Milky Way. We provide the percentages of systems that fall into each category at both Second generation and Third generation phases. The first four rows contain the numbers of systems that form discs, and how these split between different mass transfer mechanisms. We then provide the numbers of systems that do not form discs and why, and finally the number of systems that exhibit double mass transfer.}
\label{table:results}
  \begin{centering}
     \begin{tabular}{lcc}
     \hline
     \multicolumn{1}{c}{} &
     \multicolumn{1}{c}{Second generation} &
     \multicolumn{1}{c}{Third generation} \\
     \hline
     {\bf Total Viable Systems}  & 0.2635\%   & 0.1266\% \\ 
     \hline
     BHW only                    & 0.0076\%   & 0.0126\% \\ 
     WRLOF only                  & 0.0055\%   & 0.0012\% \\ 
     Both WRLOF and BHW          & 0.2497\%   & 0.1128\% \\ 
     \hline
     {\bf Total Rejected Systems}& 99.7365\%  & 99.8734\% \\ 
     \hline
     Below 20\,AU Separation     & 3.0900\%   & 1.3409\% \\
     Above 3000\,AU Separation   & 0.6740\%   & 0.3326\% \\
     Below $0.7M_{\odot}$         & 92.9826\%  & 96.8675\% \\
     Above $7M_{\odot}$           & 0.0296\%   & 0.0056\% \\
     Stellar ages too close      & 0.0029\%   & 0.0029\% \\
     Not evolved yet             & 2.1047\%   & 0.9390\% \\
     RLOF occurs                 & 0.0190\%   & 0.0079\% \\
     Disc does not form          & 0.8337\%   & 0.3770\% \\
     \hline
     {\bf Double mass transfer}  & 0.0555\%   & 0.0555\% \\
     \hline
     \end{tabular}
  \end{centering}
\end{table}

\begin{figure}
  \centering
  \includegraphics[width=\columnwidth]{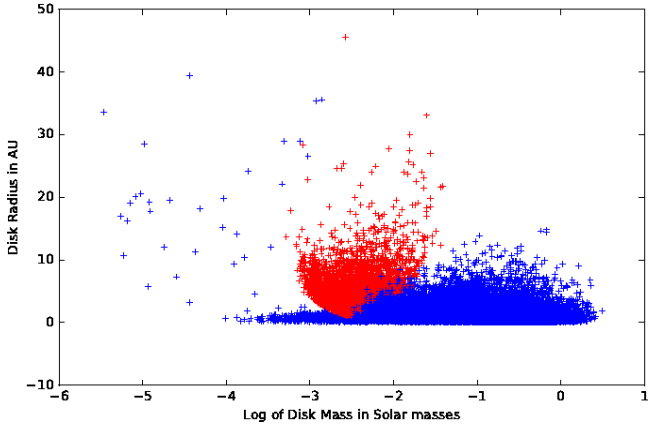}
  \caption{The disc mass plotted against the disc radius for the discs formed within our binary systems. The two mass transfer regimes are marked by red (BHW) and blue (WRLOF). The most massive discs are formed by WRLOF (which typically has a higher capture fraction), while the majority of the discs with a larger radial extent are formed in wider binaries (where BHW is more prevalent).}
  \label{fig:2}
\end{figure}

\begin{figure}
  \centering
  \includegraphics[width=\columnwidth]{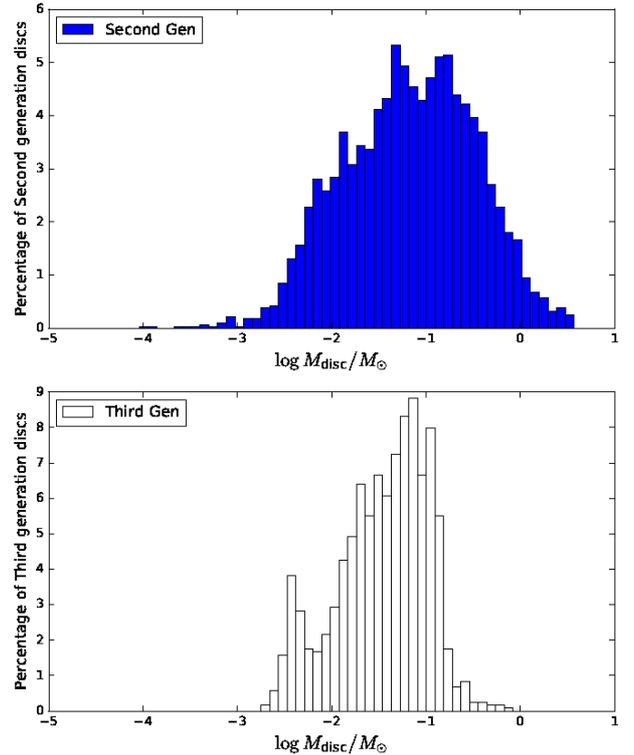}
  \caption{Histograms of the disc masses for Second generation (top panel) and Third generation (bottom panel) discs. The range of disc masses is similar in both cases, peaking around $0.1M_\odot$, but the most massive discs occur in the Second generation phase.}
  \label{fig:3}
\end{figure}

\begin{figure}
  \centering
  \includegraphics[width=\columnwidth]{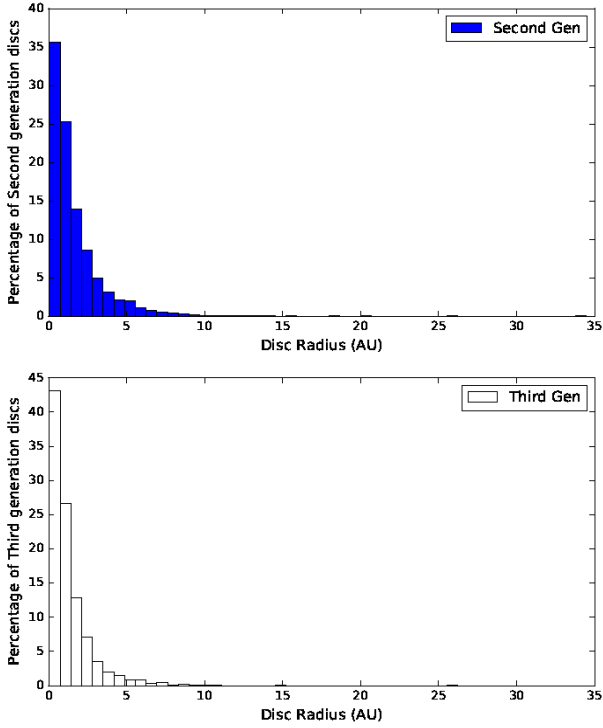}
  \caption{A histogram of the disc radii. The majority of discs which are formed are compact, with radii $\lesssim 5$\,AU.}
  \label{fig:4}
\end{figure}

In Fig.~\ref{fig:2} we show the estimates of the disc mass and radius for each of the systems that form discs. We note that for the `disc radius' we use the tidal truncation radius (\ref{eq:diskrad}), and the `disc mass' is simply the total captured mass, and so these numbers need to be understood in their context. Over time we would expect the discs to spread viscously increasing the disc outer radius from the initial circularisation radius to the tidal truncation radius within the binary. Although we note that while the discs are low-mass, they may also be cold and the conditions for e.g. the MRI to be active would need to be checked to calculate the timescale on which the disc accretes/spreads. Similarly over time the disc mass may decrease, if for example material is accreted on to the central object or lost in photoevaporative winds. In Fig.~\ref{fig:2} we see that the BHW discs are on average less massive and larger in radius than the WRLOF discs, although there exists a tail of low mass WRLOF discs. In general the WRLOF discs are compact, as expected, but due to the large mass transfer rates and high capture fractions they can be quite massive. The WRLOF regime also produces massive discs in highly eccentric systems as the capture fraction depends on the ratio of the Roche lobe filled by the donor star, and highly eccentric systems are more likely to come close enough at pericentre to increase the capture fraction and transfer a large amount of mass. However, these highly eccentric systems tend to produce smaller disc radii. Figs.~\ref{fig:3} \& \ref{fig:4} separate the data from Fig.~\ref{fig:2} into histograms of the final mass and radius. A large fraction of the discs are a few AU or less in size which may not provide the most efficient environment for planet formation. However, we note that small disc sizes are expected in the pulsar systems which have been observed to host planets which most likely formed from a fallback disc \citep[see e.g.][]{martin_livio_2016}. Future calculations of the propensity of planet formation in compact, massive discs will be required to accurately determine the number of Second and Third generation systems which host planets.

\begin{figure}
  \centering
  \includegraphics[width=\columnwidth]{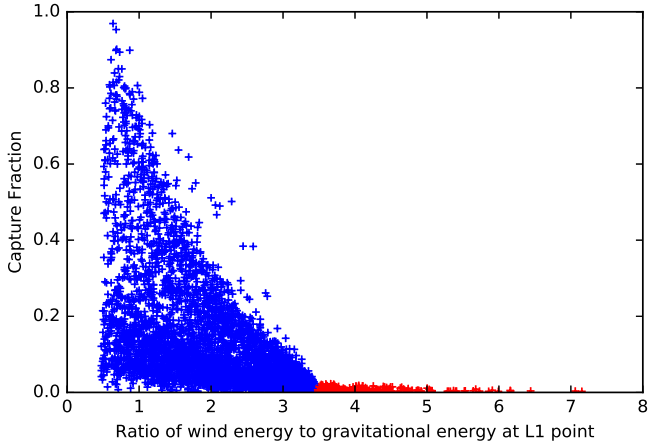}
  \caption{The capture fraction for each system (see Section~\ref{sec:MTE}) plotted against the ratio of kinetic energy of the wind at the L1 point to the gravitational potential energy at the L1 point. The blue crosses represent systems in the WRLOF regime and the red crosses represent systems in the BHW regime.}
  \label{fig:5}
\end{figure}

\begin{figure}
  \centering
  \includegraphics[width=\columnwidth]{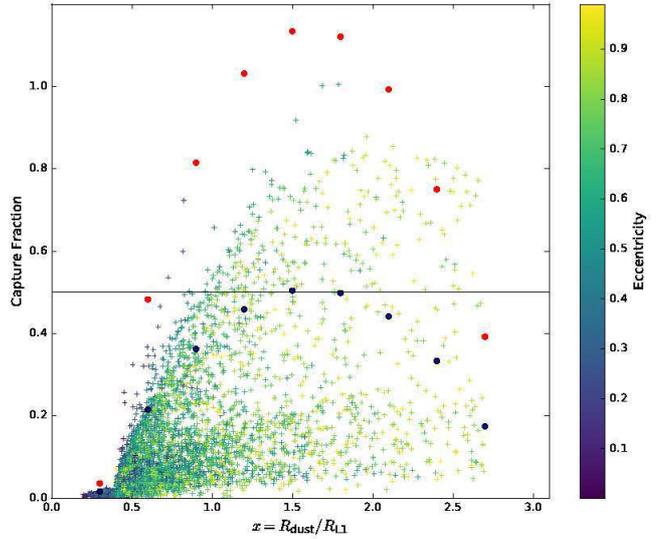}
  \caption{The capture fraction (see Section~\ref{sec:MTE}) as a function of the ratio of dust sublimation radius to Roche lobe radius, $x = R_{\rm dust}/R_{L1}$. The black line marks a capture fraction of 0.5 which is the maximum value used by \protect\cite{abate_pols_2013}. The WRLOF capture fraction for a mass ratio $q=0.9$ is marked with red filled circles. The blue filled circles denote the WRLOF capture fraction without the mass ratio dependence, i.e. the term in square brackets of (\ref{eq:wrlof}). A zoom-in at small values of $x$ can be seen in Fig.~\ref{fig:7}.}
  \label{fig:6}
\end{figure}

\begin{figure}
  \centering
  \includegraphics[width=\columnwidth]{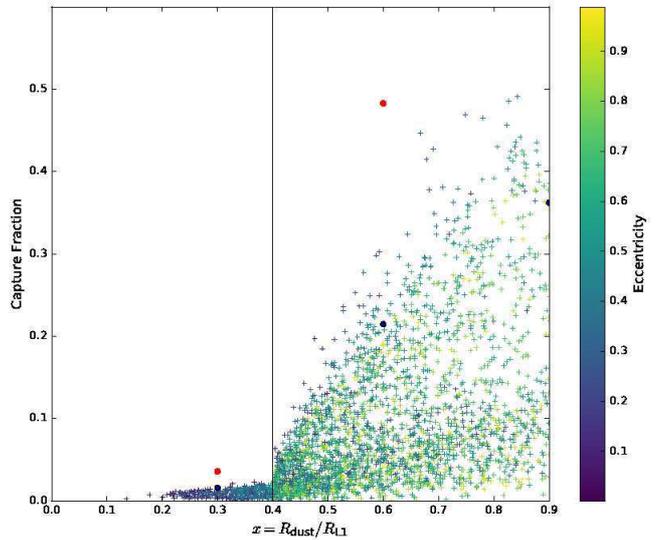}
  \caption{This is a zoomed in version of Fig.~\ref{fig:6} focusing on the change between the BHW and WRLOF regimes, marked by the black line.}
  \label{fig:7}
\end{figure}

In Fig.~\ref{fig:5} we plot the capture fraction against the ratio of the wind energy to the gravitational energy at the edge of the Roche lobe of the mass losing star. This plot shows the main difference between WRLOF and BHW. The slow winds in the WRLOF regime have only just enough energy to escape, and thus the capture fraction is high, while the BHW has higher wind velocities and thus correspondingly lower capture fractions. In Fig.~\ref{fig:6} we plot in more detail the capture fraction for the WRLOF mechanism. The critical parameter is the fraction of the Roche lobe radius filled by the dust sublimation radius of the wind of the donor star (the parameter $x$ defined in eq.~\ref{eq:x}). The WRLOF capture fraction has a quadratic dependence on $x$. This is because if the surface of the donor star is close to completely filling the Roche lobe the dust sublimation point where a slow wind is created will be too close to the Roche lobe radius and the wind will be lost through the L2 as well as the L1 point, meaning less is captured by the companion. This implies that there is a maximum point of capture by the secondary when the dust sublimation point is around half the size of the Roche lobe. Fig.~\ref{fig:6} also shows the capture fraction calculated for equal mass ratios (the term in square brackets in eq.~\ref{eq:wrlof}) in blue and the full capture fraction equation (\ref{eq:wrlof}) which includes the mass ratio dependence in red. While the full equation has a maximum capture fraction that exceeds unity (recall that it is a fit to numerical simulations, and this prompted \citealt{abate_pols_2013} to impose a maximum capture fraction of 0.5) we find that none of the systems attain a capture fraction above 0.8, and thus we do not impose a cap. All of the systems with a high capture fraction have a high (near-equal) mass ratio which is expected from (\ref{eq:wrlof}). If we reduce (\ref{eq:wrlof}) to just the term in square brackets (i.e. without the mass-ratio dependence) the capture fractions fall below the blue points in Fig.~\ref{fig:6} which reduces the number of high mass discs in Fig.~\ref{fig:2}. Note that the data points for each system do not exactly follow the predictions in Fig.~\ref{fig:6} due to the binary eccentricity. (\ref{eq:wrlof}) assumes a circular orbit so the capture fraction remains constant around the orbit. To introduce eccentricity we calculate the capture fraction at discrete intervals throughout the orbit and averaged them. In highly eccentric systems the capture fraction varies by an order of magnitude and only a small amount of the orbit is spent at pericentre. Fig.~\ref{fig:7} is a zoomed in version of Fig.~\ref{fig:6} showing the point where the regime changes between from BHW and WRLOF, the regime change is marked with a black line and illustrates that the change between the regimes is fairly smooth. 

\begin{figure}
  \centering
  \includegraphics[width=\columnwidth]{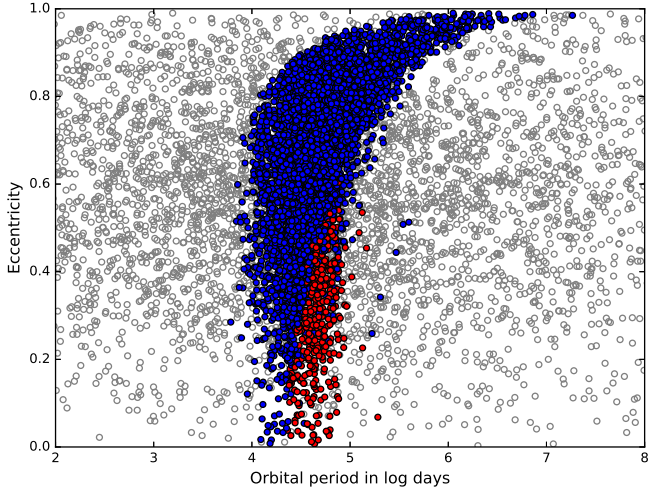}
  \caption{The distribution of WRLOF (blue) and BHW (red) systems as a function of eccentricity and orbital period. A random sample drawn from the full set of systems is included (grey circles) for comparison.}
  \label{fig:8}
\end{figure}

\begin{figure*}
  \centering
  \includegraphics[width=\textwidth]{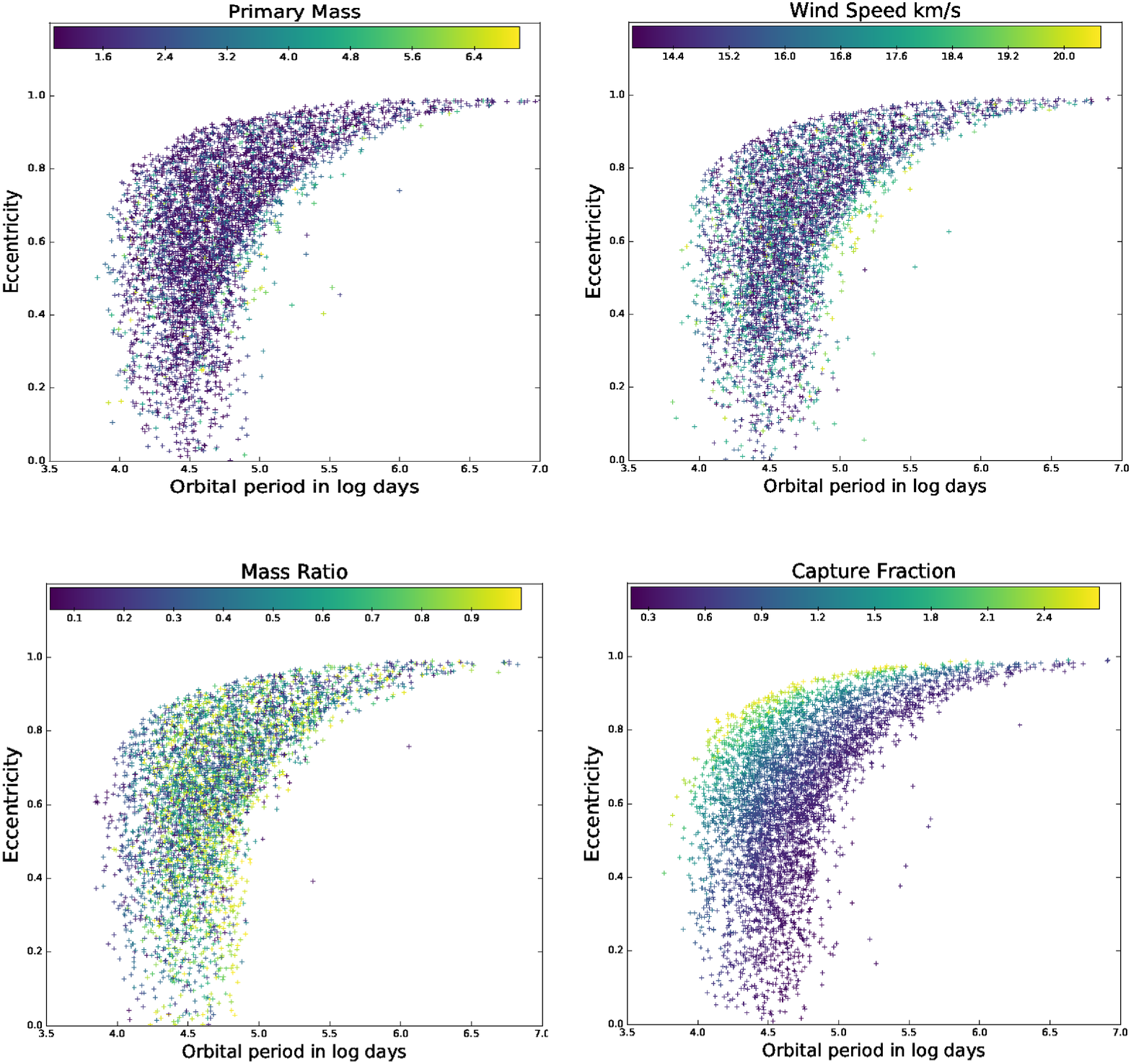}
  \caption{Plots showing the values of several parameters for each system as a function of eccentricity and orbital period. The colour of each cross (system) gives the primary mass (top left panel), wind velocity (top right), mass ratio (bottom left) and capture fraction (bottom right).}
  \label{fig:9}
\end{figure*}

Figs.~\ref{fig:8}~\&~\ref{fig:9} show the parameter space occupied by the binaries that form discs. Fig.~\ref{fig:8} compares the systems with discs (blue if WRLOF occurs anywhere during the orbit and red if BHW alone occurs) with a sample of all the systems (grey). The systems with a disc have periods of $10^{4}$ to $10^{5}$ days at low eccentricities and extends up to $10^{7}$ days at the highest eccentricities. The systems at shorter orbital periods are in the RLOF regime or end up merging. The systems at longer orbital periods are separated too widely to have appreciable mass transfer. Fig.~\ref{fig:8} shows that the BHW systems (red) are mostly low eccentricity. The systems at the extreme end of the parameter space are the high eccentricity systems which have large orbits, but come in close enough at pericentre for episodic mass transfer. These systems are unlikely to easily form standard discs, these systems might be likened to `heartbeat' binaries where the extreme orbit gravitationally or tidally affects the companion. Fig.~\ref{fig:9} shows the same systems on the same axes, but coloured by the values of different variables; namely primary mass (top left), wind speed (top right), mass ratio (bottom left) and capture fraction (bottom right). The primary mass panel show that most of the systems in the parameter space have a primary mass of $\lesssim 4M_{\odot}$, a wind speed of $\sim 15$\,km/s and a mass ratio of $\sim 0.5$. The capture fraction panel shows a correlation between eccentricity and capture fraction. This correlation is due to the difference between the capture fraction in the BHW and WRLOF regimes, the WRLOF regime has a much higher capture fraction so any binary with WRLOF at some point in its orbit will have an increased capture fraction, The curve at the top of these graphs is the point where the eccentricity is so high that the binaries enter RLOF.

\begin{figure}
  \centering
  \includegraphics[width=\columnwidth]{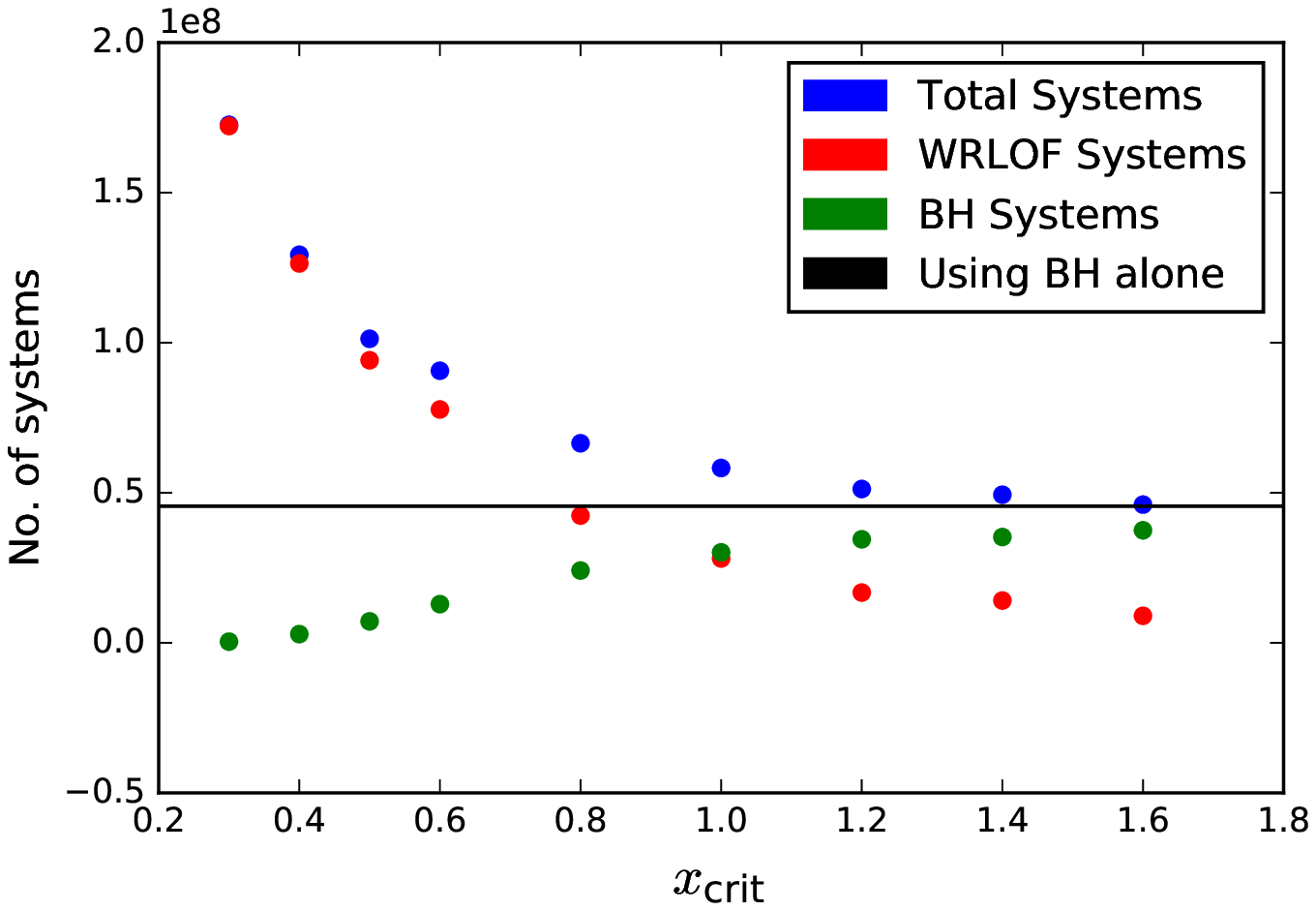}
  \caption{In this plot we show the effect of varying the critical value, $x_{\rm crit}$, of $x = R_{\rm dust}/R_{L1}$ at which the mass transfer mechanism changes from WRLOF to BHW. The default parameter is $x_{\rm crit} = 0.4$ \citep{abate_pols_2013}, and we vary this from 0.3 to 1.6. The black line denotes the number of systems classified as BHW when this is the only option. The filled circles represent the number of systems from our analysis including WRLOF for 9 population synthesis models each with a different $x_{\rm crit}$ value. The green circles represent the systems in the BHW regime, the red circles represent those in the WRLOF regime and the total (BHW+WRLOF) is given as blue circles. The inclusion of WRLOF has two main effects, the first is to increase the total number of systems which form discs, and the second is that many of the BHW systems at small $x$ are actually in the WRLOF regime.}
  \label{fig:10}
\end{figure}

The number of discs formed is highly dependent on which mass transfer regime the binary systems fall into. The mass transfer regime is chosen by a critical value of the parameter $x=R_{\rm dust}/R_{L1}$, which \cite{abate_pols_2013} set to 0.4. We have used a default value of $x_{\rm crit} = 0.4$, but we test the effect of varying $x_{\rm crit}$ in Fig.~\ref{fig:10}. As $x_{\rm crit}$ varies we see a change in the number of discs formed. As we increase the cutoff between regimes we see an overall decrease in total discs towards the value found when only BHW is considered (black line). This is as expected, as we increase the radius the donor needs to exceed to be in the WRLOF regime, and therefore the number of systems that fulfill this criteria drops significantly. The BHW disc formation condition is more difficult to achieve so as more systems fall into this regime less discs are able to be form. Therefore we can see that the value of $x_{\rm crit}$ is important in determining the numbers and properties of discs formed in these binary systems. However, varying $x_{\rm crit}$ across the possible range of values does not lead to a change in our conclusions, and leads to a change in the number of discs by a factor of order 2-3.

\begin{figure}
  \centering
  \includegraphics[width=\columnwidth]{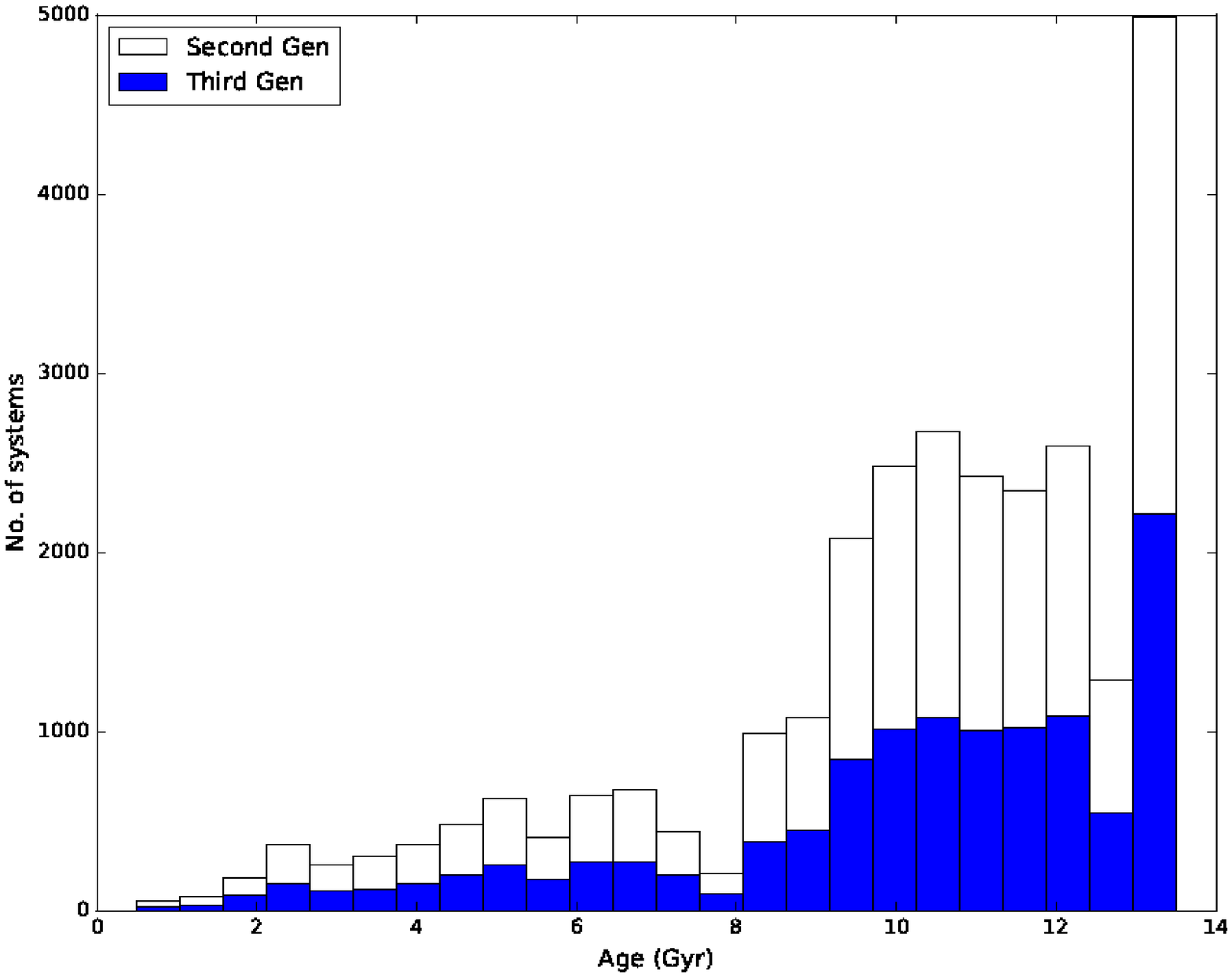}
  \caption{This plot shows the age distribution of the systems with second and third generation discs, it is similar to the age distribution of the full sample shown in Figure \ref{fig:1} but has notably less systems that are younger in age, due to the system needing to have evolved.}
  \label{fig:11}
\end{figure}

Finally, Figure \ref{fig:11} shows the age distribution of the systems with second and third generation discs. This figure shows that the disc-bearing systems favour older system ages. This result is expected as the systems will have needed time to evolve so the earlier they formed the more likely they are to have evolved by the current Milky Way age. We note that over half of the systems with discs are older than 11 Gyr which means they would likely be located in the thick disc or bulge of the Milky Way.

\section{Discussion}\label{sec:discussion}
We have used a binary population synthesis tailored to parameters of the Milky Way to explore the binary systems in which Second and Third generation discs may form. Our approach utilises the properties of binary systems determined from observations, and theoretical understanding of mass transfer in binaries to determine if these binaries allow the formation of discs through either BHW or WRLOF. 
 
For the Third generation phase we calculate the new separation of each binary system by assuming all mass lost from the initial primary is lost from the system, and thus the expansion of the binary is maximal. However, in reality some mass is retained in the system, either through accretion onto the companion or left orbiting the companion. We can consider how this affects the results of our Third generation systems by looking at the two extremes: where all mass is lost (our default) and where all of the transferred mass is retained. When we retain all the mass, implying that the binary separation does not change there is no change in the amount of Second generation discs (which are unaffected), but there is a small increase (2\%) in Third generation discs. This is because when mass loss is included some binary systems expand beyond our cutoff separation of 3000\,AU. To test the full range we also performed the population synthesis with different percentages of mass retained by the system and found little change in the overall numbers of Third generation discs.

Much of the AGB phase is still uncertain in detail as there are too few observations of AGB stars to build a complete picture. In modelling the AGB phase we have made some simplifying assumptions, such as taking all of the stars to have solar metallicity. The inclusion of varying metallicity would affect some of the parameters such as wind speed and radius of the AGB star. However most of the uncertainty in our population spread comes from the mass regimes, and we regard the differences we would get from varying the metallicity as negligible in comparison. We also assume that mass loss only occurs in the AGB phase. There is some mass loss in the RGB phase, but as most of the mass is lost in the AGB phase and this is also where we expect WRLOF to occur, the AGB phase is our area of focus. We also ignore the time dependence of the mass loss (pulsations and bursts) for simplicity.
 
We do not take into account the subsequent evolution of the transferred mass, instead we calculate the disc radii and the total mass transferred into a disc during the AGB phase. This gives an upper limit on the disc masses. The lower right hand side in Fig.~\ref{fig:2} is an area that would be unlikely to form planets. The high mass, small radii discs are formed from the systems that are close in separation with a primary near the maximum $7M_{\odot}$ limit. These systems are the ones on the cusp of RLOF, but are just within the WRLOF regime, giving a high mass transfer rate. If the discs become this massive, they are likely to be self-gravitating with masses of order the central star's mass. If they are active discs, either through self-gravity or hot enough for e.g. the MRI to be active, then the disc mass could accrete onto the central star and chemically enrich the surface. This may be a route to forming carbon or barium enhanced peculiar stars. Barium stars are often found in binaries with an orbital period of $10^{2.7-4}$\,days and have some eccentricity. \cite{moe_di-stefano_2017} suggest that 3.1\% of main sequence stars with a white dwarf companion and an orbital period $<10^{4.7}$ days are Barium enhanced stars. To check the prediction from our population, we look for solar-like stars between $0.9-1.3 M_{\odot}$ and count those that form a disc larger than $0.01M_{\odot}$ and find a percentage of 3.06\% which is remarkably close to the observed value. We also count discs in different mass ranges to see how it affects the number of chemically peculiar stars and find the percentage changes to 4\% for discs above  $0.001M_{\odot}$ and 1\% for discs above $0.1M_{\odot}$. These values are close to the observed value.

The exact process of planet formation is still debated, and so the parameters of the discs in which they form are not precisely known. Planet formation theories have also, naturally, focussed on the formation of planets in protoplanetary discs around young stars. The discs formed in binary systems at Second and Third generation have different properties. Most notably these discs have a much higher dust content which affects the opacity and temperature of the discs. The metallicity of these discs would also be much higher than typical protoplanetary discs as the mass is transferred from the enriched outer atmosphere of the AGB stars. We can assume that for planets to form we need the disc to extend beyond the snow line for solids to form efficiently. In protoplanetary discs, depending on the flux from the central star and the disc thermodynamics, the snow line is at $\sim 2$\,AU \citep{lecar_podolak_2006}. \cite{nixon_king_2018} have recently argued that the efficient formation of planets is aided by disc self-gravity causing gaseous spiral arms that act as dust traps to efficiently grow planetesimals on short timescales \citep[see also][]{Rice:2004aa,Rice:2006aa}. In this case we would need the disc mass to be above $\sim 0.02 M_\odot$. With these basic limits we can estimate the fraction of discs formed in the binaries in our population with the potential to form planets. We find that 20\% of Second generation and 3.8\% of Third generation discs satisfy these constraints. For the Milky Way, which we expect to host $\sim 50$ billion binary systems, we therefore estimate that 30 million systems (0.06\%) have the potential to form planets during the Second generation, and 3.5 million (0.007\%) during the Third generation. \\

In  our initial assumptions we do not include the prospect of first generation planets that may already exist in a binary system. First generation planets or planetesimals may act as 'seeds' for planet formation and growth \cite[see][for further reading]{peret_2010, peret_2011}. These 'seeds' may assist the growth of the second generation planet or change the properties of the existing planet with the accretion of metal rich material. The addition of first generation planets or planetesimals into the second generation formation scenario could potentially enhance the fraction of discs that form second and third generation planets. Due to the additional factors that would be involved in 'seeding' second generation planets we have not included this in our current work but it should be noted as it is a mechanism that could enhance planet formation in these discs. \\

The majority of binaries that have been catalogued to date are either shorter period eclipsing binaries or far enough apart to visually resolve. However, the binaries that we estimate are promising for disc, and potentially planet, formation are in between these two. The catalogues for these systems have a much lower completeness factor. Third generation systems with a white dwarf component are especially incomplete as the white dwarf would be much dimmer than the main sequence or giant star and they are at a separation that makes them difficult to resolve visually or spectrally. We expect that the imminent Gaia data release, which will give proper motion, radial velocity, astrometric and photometric data for more than a billion stars, will help give a more complete data set. The new data release will hopefully find enough of these intermediate separation binaries so we can more efficiently search for systems that have the conditions we suspect would form second and Third generation discs. Protoplanetary discs formed around single stars last for around 1-10 Myr and discs are expected to have significantly shorter lifetimes of 0.3-5 Myr in binaries \citep{williams_cieza_2011}, we can assume that second and Third generation discs will have a similar lifetime so the chances of observing them are small, requiring a relatively large sample.

\section{Conclusions}\label{sec:con}
We have performed a population synthesis of binary systems to determine the number of systems which form second and Third generation discs. Our population is tailored to the Milky Way, employing for example the observed star formation history, binary fractions and orbital period and eccentricity distributions. We have also used mass transfer equations callibrated by numerical simulations and corroborated by observational data of AGB systems. Of the three main mass transfer mechanisms (RLOF, WRLOF and BHW) we have focussed on WRLOF and BHW, as RLOF discs are too small and hot to be potential sites of planet formation \citep{perets_kenyon_2013}. We have also focussed on S-type planets (those in circumprimary or circumsecondary discs). This could be extended in the future to include P-type planets (circumbinary discs) if some predictions for the capture fraction of material into circumbinary discs were available. For the systems where disc formation is predicted, we have calculated disc radii and masses. The disc masses are upper limits as mass may be removed by e.g. accretion on to the central star or lost to photoevapourative winds. Future calculations, including the time-dependence, of these discs would be required to determine the precise disc conditions. However, we have shown that significant numbers of systems have the potential to create discs with masses high enough that planetary systems can be expected to form \citep[e.g. the Maximum Mass Solar Nebula;][]{nixon_king_2018}. We provide the main results of our analysis below:
\begin{itemize}
\item A small percentage of systems ($\sim 0.1-1$\%) will transfer mass and form discs. Given the number of binary systems in our galaxy this is a significant population.
\item High eccentricity systems are more likely to have mass transfer and form discs. However, such eccentric binary systems may be an inhospitable environment for planet formation, and could instead (or as well as) lead to metal enrichment of the accreting stars.
\item Lower wind speeds from the donor star will increase the number of BHW discs that are formed, which typically have a larger radius than the WRLOF discs.
\item Over 75\% of systems that create discs exhibit both WRLOF and BHW rather than falling into a single mass transfer regime.
\item The mass transfer rates in some of these systems are high enough that they may cause different phenomena such as metallicity enhancement or nova.
\item The number of discs that could potentially enrich the companion star is $~3\%$ of the binary systems in our population, which is close to the observed value of 3.1\%.
\item Second and Third generation discs have a higher dust and metal content compared to (First generation) protostellar/protoplanetary discs.
\item Finally, we find that 0.27\% of binary systems will host Second generation discs and 0.13\% of systems will host Third generation discs. For the Milky Way, this translates into $\sim 130$ million and $\sim 90$ million systems with second and Third generation discs respectively. Of these we estimate approximately 20\% and 3.8\% of second and third generation planets have enough mass to form a planetary system of comparable size to the Solar System.
\end{itemize}

\section*{Acknowledgements}
We thank our reviewer, Hagai Perets, for his constructive and thoughtful comments on the manuscript.
CJN is supported by the Science and Technology Facilities Council (grant number ST/M005917/1). MAH is funded by the Science and Technology Facilities Council.
  
\bibliographystyle{mnras}
\bibliography{MahPapers}

\bsp
\label{lastpage}
\end{document}